\pdfoutput=1

\documentclass[twocolumn]{RAA}
\usepackage{graphicx,times}             %for PS/EPS graphics inclusion, new
\usepackage{natbib}
\usepackage{amssymb,amsmath}
\usepackage{changes}

\bibpunct{
(}{)}{;}{a}{}{,}

\usepackage[pagebackref=true]{hyperref}

\newcommand{\HIFASTGAIN}{\texttt{HiFAST.FluxGain} }
\def\HIFAST{\texttt{HiFAST} }
\def\HI{H\,{\small{I}} }

\begin{document}

\title{\HIFAST: An \HI Data Calibration and Imaging Pipeline for FAST
}
\subtitle{II. Flux Density Calibration}

   \volnopage{Vol.0 (20xx) No.0, 000--000}      %%preserved for Editor. DOn't remove!
   \setcounter{page}{1}          %%starting page, preserved for Editor. DOn't remove!

   \author{Ziming Liu \inst{1,2}%(刘孜铭) 
   \and Jie Wang \inst{1,2}
   \and Yingjie Jing \inst{1} 
   \and Zhi-Yu Zhang \inst{3,4}
   \and Chen Xu\inst{1,2}
   \and Tiantian Liang\inst{1,2}
   \and Qingze Chen \inst{1,2}  
   \and Ningyu Tang \inst{5}
   \and Qingliang Yang \inst{1}
   }
   %% Put your Chinese name in "( )" if you like. Note to open line 11 "\usepackage[UTF8]{ctex}"
   

   \institute{
    National Astronomical Observatories, Chinese Academy of Sciences,
   Beijing 100101, China; {\it jie.wang@nao.cas.cn}\\
        \and
    School of Astronomy and Space Science, University of Chinese Academy of Sciences, Beijing 100049, China \\
        \and
    School of Astronomy and Space Science, Nanjing University, Nanjing 210023, China\\
        \and
    Key Laboratory of Modern Astronomy and Astrophysics, Nanjing University, Nanjing 210023, China\\
        \and
    Department of Physics, Anhui Normal University, Wuhu, Anhui 241002, China\\
\vs\no
   {\small Received 20xx month day; accepted 20xx month day}}

\abstract{Accurate flux density calibration is essential for precise analysis and interpretation of observations across different observation modes and instruments. In this research, we first introduce the flux calibration model incorporated in \HIFAST pipeline, designed for processing \HI 21-cm spectra. Furthermore, we investigate different calibration techniques and assess the dependence of the gain parameter on the time and environmental factors. A comparison is carried out in various observation modes (e.g. tracking and scanning modes) to determine the flux density gain ($G$), revealing insignificant discrepancies in $G$ among different methods. Long-term monitoring data shows a linear correlation between $G$ and atmospheric temperature. After subtracting the $G$--Temperature dependence, the dispersion of $G$ is reduced to $<$3\% over a one-year time scale. The stability of the receiver response of FAST is considered sufficient to facilitate \HI observations that can accommodate a moderate error in flux calibration (e.g., $>\sim5\%$) when utilizing a constant $G$ for calibration purposes. Our study will serve as a useful addition to the results provided by \citet{jiang2020fundamental}. Detailed measurement of $G$ for the 19 beams of FAST, covering the frequency range 1000 MHz -- 1500 MHz can be found on the \HIFAST homepage: \url{https://hifast.readthedocs.io/fluxgain}.
\keywords{telescopes---techniques: image processing---methods: observational---radio continuum: galaxies---cosmology: observations}
}
   \authorrunning{Z.M. Liu et al.}            %author_head in even pages
   \titlerunning{Flux density calibration in FAST observation }  % title_head in odd pages

   \maketitle
%% The author head (on even pages) and the title head (on odd pages) will be
%% automatically extracted from \author{} and \title{}. Whenever the title is too long,
%% you will be asked to supply a shorter one by inserting either \authorrunning{} or
%% \titlerunning{} before \maketitle. Anyway, you can specify your own heads.
%%
%%
%% Note: In the following text body of your manuscript, please note several differences from
%%       other major journals:
%% (1) \subsection{Please Capitalize the First Letter of Each Notional Word in Subsection Title}
%% (2) Please Capitalize the First Letter of Each Notional Word in all tables' captions
%
%________________________________________________ sections below
%
\section{introduction} %% first-level sections will be auto-capitalized
\label{sect:Introduction}

Observational data acquired from radio telescopes are often subject to a multitude of origins of biases, including weather conditions, radio frequency interference (RFI), hardware drift, deformation of the telescope panels, etc. Hence, calibrating the acquired data is imperative to derive reliable results. Well-established methods utilize robust and stable radio sources as calibrators, with extensive literature documenting their efficacy \citep{kuzmin2013radioastronomical, baars1973measurement, baars1977absolute, boothroyd2011accurate, salter2000notes}. Fluctuations in the hardware configuration of a system can induce variations in the gain of a radio telescope, necessitating regular calibration of the electronic system \citep{maddalena2004methods}. The choice of calibration method is contingent upon several factors, including the observing strategy (e.g., observation modes, data sampling time), frequency range, and telescope performance. Hence, selecting the appropriate calibration method is pivotal for obtaining accurate results in radio astronomy observations.

Among the various calibration processes, precise flux density calibration is particularly crucial, especially for highly sensitive single-dish telescopes. Single-dish telescopes serve as indispensable tools for probing the large-scale, low-density distribution of gas \citep{van1997obtaining}. The Five-hundred-meter Aperture Spherical radio Telescope (FAST) completed its commissioning phase in 2019 \citep{nan2006five,nan2011five,jiang2019commissioning}. Featuring a 19-beam receiver design and unparalleled sensitivity, FAST stands as an optimal instrument for conducting large-scale \HI surveys \citep{li2018fast, duffy2008galaxy}. Several forecasts regarding FAST \HI spectral research have been articulated, including its capability to spearhead the next generation of \HI blind sky surveys aimed at detecting high-redshift galaxies and other faint radio sources \citep{duffy2008galaxy}. Ongoing initiatives such as the Commensal Radio Astronomy FAST Survey (CRAFTS, \citet{li2018fast}) are underway, with \citet{zhang2019status} providing insights into the survey's extragalactic \HI sensitivity and projecting FAST's future survey capabilities.
 
To realize the scientific objectives outlined above, precise flux density calibration is imperative for the sky surveys and observations conducted using FAST. However, the existing data open-source data-processing pipelines for FAST lack a systematic flux density calibration procedure. Therefore, we have developed and integrated a module into the \HIFAST pipeline specifically designed for flux density calibration of observational data, followed by an evaluation of its performance. 

Several observation modes are offered to observe calibrators by FAST, including OnOff, MultiBeamCalibration, and Drift. Selecting the most suitable calibration mode for a given observation is crucial. Furthermore, the stability and variability of the gain depend on both the frequency and the directionality of the telescope. As FAST utilizes a transformable paraboloid surface for observation, the mirror undergoes deformation at varying elevations and azimuth angles, potentially leading to fluctuations in observation gain throughout the observation process. The stability of the FAST electronic system meets the accuracy requirement ($<1\%$) for observations with calibrators \citep{jiang2020fundamental}. Despite the importance of calibration, its implementation is often constrained by limited observation time and the availability of calibrators. Notably, calibrating each of the 19 beams individually consumes a significant amount of observation time. Thus, in scenarios where calibration is unfeasible, a fixed $G$ value is adopted for flux density calibration. \citet{jiang2020fundamental} provided a version of the FAST $G$ table (hereafter $G_{J2020}$) with a fitting function of zenith angle ($\theta_{ZA}$) at specific frequencies. It is anticipated that observation modes and external environmental factors might also affect $G$. To date, the accuracy of flux calibration across various observation modes has not been thoroughly investigated. Furthermore, there has been no examination of the long-term stability of the gain parameter of FAST. This study aims to investigate different calibration techniques and assess the accuracy of the gain parameter over an extended duration. This might serve as an addition to $G_{J2020}$.

In this paper, the calibration selection criteria and observation strategy are delineated in Section~\ref{sect:observation}. Following that, the data reduction procedures for various observation modes used in calibration are expounded upon in Section~\ref{sect:data}. Section~\ref{sect:randd} presents a comparative analysis of the $G$ values obtained from different observation modes and examines the gain curve across the 19 beams. Additionally, the section delves into the longitudinal variations of $G$ observed during extended observations. After this, Section~\ref{sect: disc} encompasses in-depth discussions on the outcomes. Finally, Section~\ref{sect:Conclusions} offers a succinct summary.

\section{Calibration data}
\label{sect:observation}
In this section, we will first introduce the choice of calibrator sources, then the observation modes often used for calibration. Finally, the introduction to the calibration data used in this paper will be presented. 

\subsection{The choice of flux calibrators }
\label{sect:calibrator}
The choice of absolute flux density calibrators is based on four criteria: (i) the source should have been regularly monitored to ensure that their L-band fluxes do not vary with time, or within a limited variation range \citep{perley2017accurate}; (ii) spatial confusion in calibration should be avoided, meaning that there should be no other continuum sources listed in the NVSS catalog provided by \citet{condon1998nrao} close to the calibrator; (iii) the coordinates of the calibrator should be as close as possible to the target sources; and (iv) and the target should be a point source, with its angular extent much less ($<<$ 10\% or less) than the beam size. 

For our test sky coverage around the M31 galaxy, 3C48 (also known as J0137+3309) and 3C286 (also known as J1331+3030), were chosen as calibrators. These sources have been extensively observed by VLA, VLBA, MERLIN, and EVN \citep{an2010kinematics} and are usually used as calibrators in observations with VLA and other telescopes due to their stability flux and small angular size \citep{perley2013accurate}. The flux densities of 3C48 and 3C286 can be accurately fitted by a polynomial function:

\begin{equation}\label{fluxfreq}
  \log[S_c(\nu)]= a_0 + a_1 \log(\nu) + a_2[\log(\nu)]^2 + a_3[\log(\nu)]^3,
\end{equation}

where $S$ denotes the ﬂux density in Jy at the frequency $\nu$ in GHz. The coefficients used are $a_0= 1.3253$, $a_1= -0.7553$, $a_2= -0.1914$, and $a_3= 0.0498$ for 3C48, and $a_0= 1.2481$, $a_1= -0.4507$, $a_2= -0.1798$, and $a_3= 0.0357$ for 3C286 from \citet{perley2017accurate}.

In the frequency range of 1000-1500 MHz, the flux density of 3C48 exhibits negligible polarization, with a polarization difference of 0.3\% at 1050 MHz and 0.5\% at 1450 MHz as reported by \citet{perley2013integrated}. In contrast, 3C286 serves as a reliable polarization calibrator due to its stable polarization degrees, angles, and total flux density within this frequency range \citep{perley2013accurate}.

The two calibrator sources we selected have very similar values of zenith angles in culmination time: $\sim 8^\circ$  for 3C48, and $\sim 5^{\circ}$ for 3C286. The difference is less than $3^\circ$. The observations of the calibrators are always processed near their culmination time, and the changes on zenith angles during observation are always less than $1.2^{\circ}$ for any DriftWithAngle, OnOff, MultibeamCalibration, or MultibeamOTF mode observations. According to $G_{J2020}$, this difference has minimal impact on the gain. So no additional correction on the $G$ during these observations is needed. Anyway, for all mode observation data, we followed the correction table provided in $G_{J2020}$ and corrected the gain parameter in all of our samples to the zenith angle of $10.15^{\circ}$.

\subsection{The modes of calibrations}
\label{sect:observationmode}

\begin{figure}
\centering
\includegraphics[width=0.5\textwidth, angle=0]{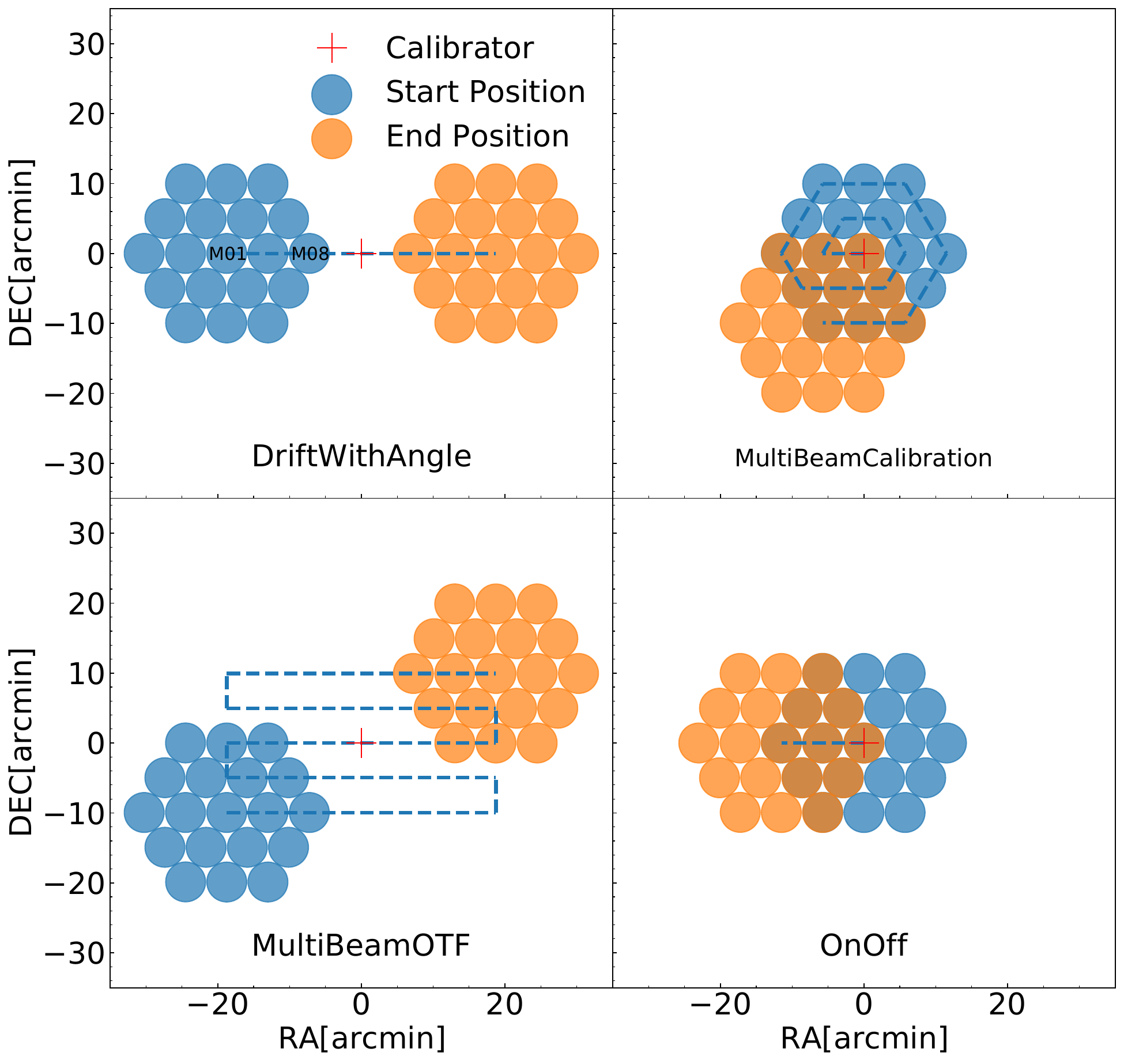}
\caption{The schematic of observation modes used in calibration: DriftWithAngle (left-top), MultiBeamOTF (left-bottom), MultiBeamCalibration (right-top), OnOff (right-bottom). The geometry of 19 beams is shown in blue dot (start position) and orange dot (end position), the red cross is the position of the calibrator, and the blue dashed lines show the desired traces of Beam 1. The relative positions of M01 and M08 are also marked on the first panel. }
\label{FigObsModes}
\end{figure}

The accuracy of the calibration highly relies on the calibration mode, i.e., the observing strategy of the calibrators. The calibration process first involves a pointing calibration. If the calibrator source is significantly smaller than the beam size, such that it can be treated as a point source, the flux variation can be effectively represented by the beam response of the telescope\citep{2002ASPC..278..293O}. The beam response of the FAST receivers exhibits high circular symmetry and can be described using a 2-dimensional Gaussian model. Then the flux density of the calibrators can be determined by fitting a Gaussian function to the observed response changes in a 1-dimensional scan, as demonstrated in \citet{jiang2020fundamental} and \citet{xi2022fast}. 

FAST provides various observation modes for calibration, and we examined four of them: DriftWithAngle, MultiBeamOTF, MultiBeamCalibration, and OnOff. Figure~\ref{FigObsModes} provides a brief illustration of these four observation modes. The desired trajectory of Beam 1 is denoted by blue dashed lines, while the configuration of the 19 beams is represented by blue and orange dots. In the following, we describe the four observing modes of calibrations adopted by FAST:

 \textbf{OnOff.} This mode enables a rapid positional switch between the source and reference positions with a slewing time of 30 seconds. Both positions are observed with the same beam and with the same duration. For multiple beam receivers, such as the L-band 19-beam receiver on board FAST, the separation between positions can be optimized for the simultaneous measurement of two beams. In our test, beams 1 (the center beam) and 8 (one of the outermost beams) were chosen for calibration, and the duration time was set to 60 seconds to reduce the error caused by the telescope's morphology instability at the start of the observation. A high noise diode signal with a synchronized period of 2.01326592 seconds was injected during the observation, with a signal strength of approximately 11 K. In total, it took 150 seconds to calibrate the two beams, including 60 seconds of tracking on the source, 60 seconds of tracking off the source, and 30 seconds of slewing time.

 \textbf{MultiBeamCalibration.} This mode is able to calibrate all 19 beams automatically by tracking the source with each beam in succession, following the order: 1-1-2-3-4-5-
6-7-19-8-9-10-11-12-13-14-15-16-17-18. Each beam requires 40 seconds to switch, and the same amount of time is needed to track the beam and set the noise diodes. This process takes a total of 1960 seconds to complete.
 
 \textbf{DriftWithAngle.} In this mode, the receiver was rotated at an angle of $0^\circ$ and the calibrator was scanned at a rate of $15^{\prime\prime}\mathrm{s}^{-1}$ with scan lines of length $67^\prime.5$. This setup enabled the centers of beams 1, 2, 5, 8, and 14 to pass the calibrator in one scan with a sufficient baseline on either side. The entire observation took 270 seconds to calibrate the five beams and the high noise diode signal was injected only at the start and end of the scanning lines with a duration of 1.00663296 seconds.
 
 \textbf{MultiBeamOTF.} Similar to the DriftWithAngle mode, this mode repeats five times with a rotation angle of $0^\circ$ to calibrate all 19 beams. Scanning lines are spaced at an interval of $4^\prime.97$, and the same noise diode injection technique as the DriftWithAngle method is used. The total time needed for one MultiBeamOTF observation to calibrate all 19 beams is 1260 seconds.
 
The calibrator observations for all modes were recorded using the spectral backend with a sampling rate of 0.201326592 seconds and a channel width of 7.62939 kHz.

\subsection{Pointing accuracy.}

\label{sect:pointing}
\begin{figure}
\centering
\includegraphics[width=0.5\textwidth, angle=0]{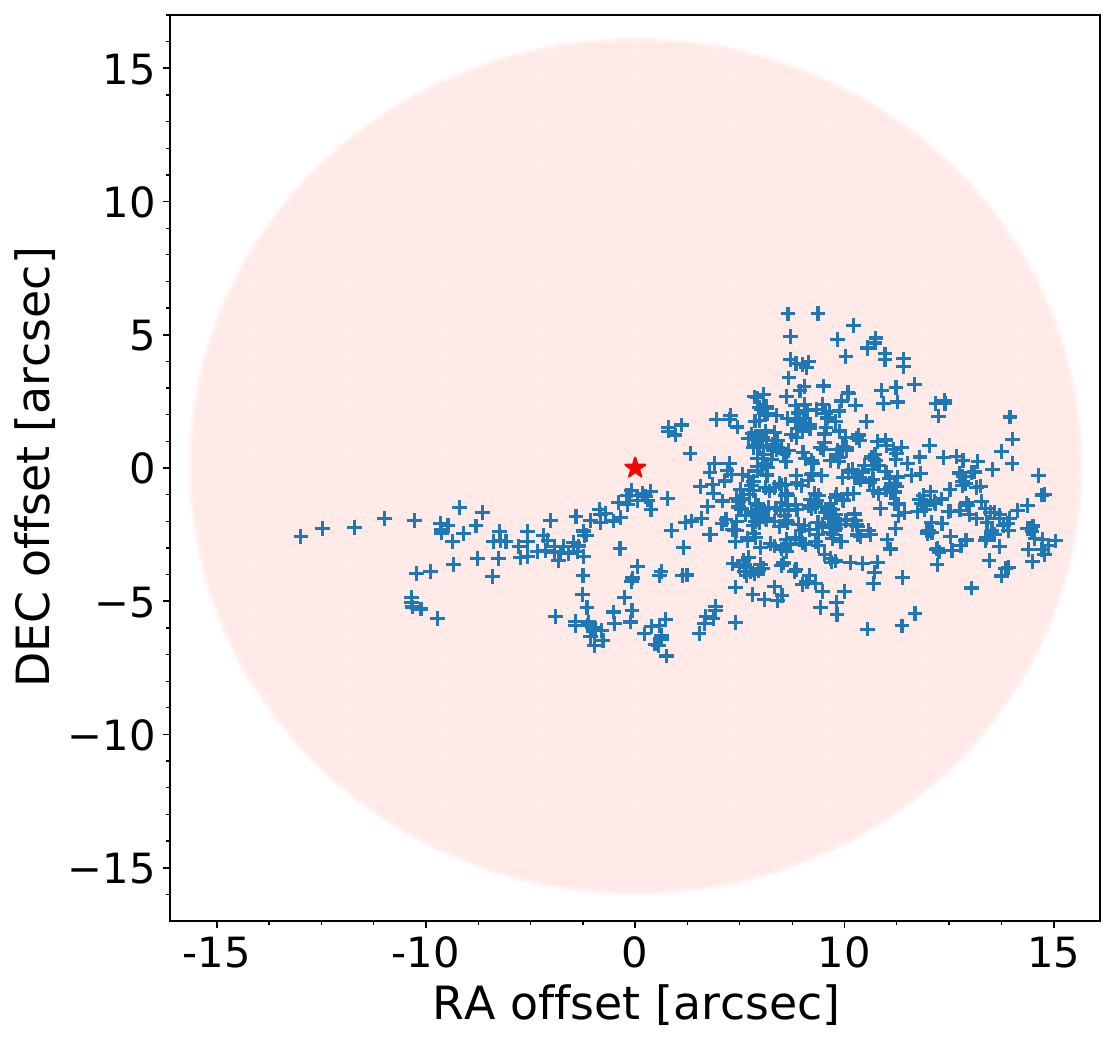}
\caption{Relative position offset measured in OnOff mode observations. Blue plus points show the coordinates of sampling points in the source area observed on 2021 September 9, distinguished by the observation time record by FAST. The red shadow area shows the region that is included in $16^{\prime\prime}$ of 3C48. The red triangle presents the center of the calibrator, 3C48.}
\label{FigTrack}
\end{figure}

\begin{figure}
\centering
\includegraphics[width=0.5\textwidth, angle=0]{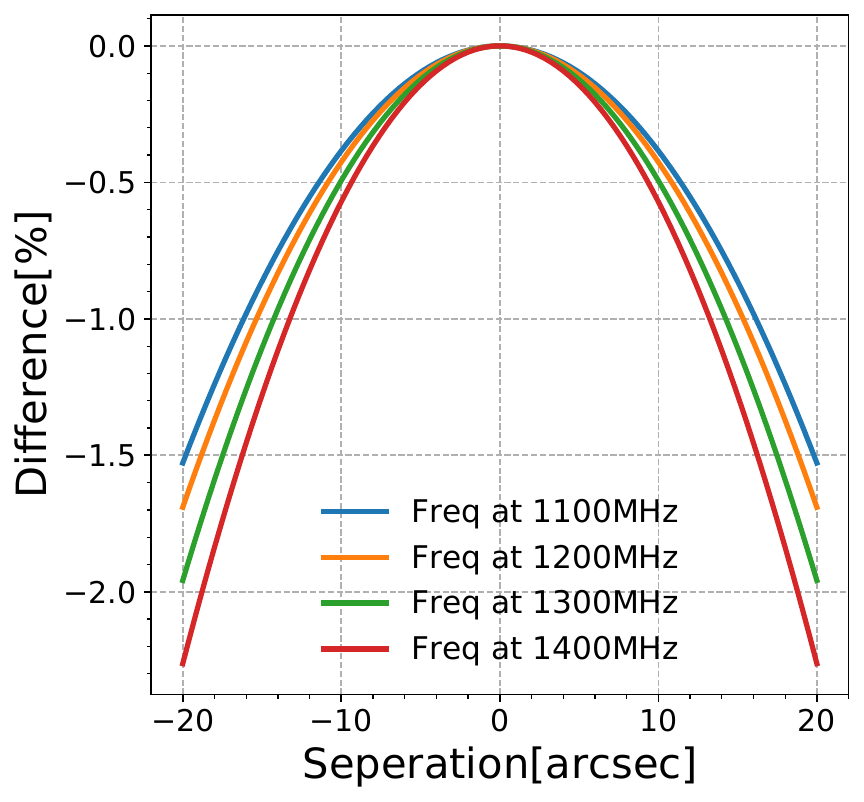}
\caption{Gaussian function fitting of the 1--dimensional scan of M01 at a different frequency. The $x$ --axis is the RA offset of sample points with the calibrator or beam center, and the $y$--axis shows the difference of detected power in percentage.}
\label{fig:scan-prof}
\end{figure}

We should be aware that the telescope's pointing accuracy can also influence the measurement of the parameter $G$. The pointing error of FAST included errors from both the surface precision of the reflector system and the accuracy of the feed support system. The measurement error of the reflector system is primarily influenced by spherical aberration, deviation in measurement shafting, atmospheric refraction, and the accuracy of the measurement reference net and the measurement errors of the feed support system come mainly from the error of the measurement reference net, the target error of the total station, the random error of the total station, and the time delay\citep{jiang2019commissioning}. Raster scan observations have been conducted to measure the pointing errors of FAST through raster scan observations\citep{jiang2020fundamental}. \citet{jiang2020fundamental} used a $2-D$ Gaussian model to model the pointing errors. This model is suitable because the central beam does not exhibit significant beam ellipticity or coma features. The surface precision of the FAST reflector system is expected to reach an RMS of $8^{\prime\prime}$ \citep{jiang2019commissioning}. \citet{jiang2020fundamental} confirmed that the pointing error of the FAST beams is smaller than $16^{\prime\prime}$, with an RMS of $7^{\prime\prime}.9$. An example of the coordinate distribution of the sample points of the OnOff observation on 2021 September 9, is shown in Figure \ref{FigTrack}. The red shadowed area indicates the filtered range of $16^{\prime\prime}$. The pointing remains within the telescope's pointing error range ($16^{\prime\prime}$). However, occasionally, a few points may exceed this range. To reduce those effects, we filtered out these outlier sample points during data processing. Through the analysis of scanning calibration conducted at various frequencies, the fluctuation of gain values near the beam's center with pointing offset has been investigated. This examination reveals the impact of pointing accuracy on calibration observations in 1-dimension, as depicted in Figure~\ref{fig:scan-prof}. We found that the flux attenuation caused by the FAST pointing error was always less than 1\%, so this error was not taken into account in our subsequent analysis.

The flux bias of pointing RMS errors, which occur when utilizing a single dish telescope to observe a point-like source, is corrected by a correction coefficient $f_G$ derived from the following equation \citep{2010ApJ...711..757P}:

\begin{align}
f_G(\sigma,\nu)= 1 + 8\ln{2}(\frac{\sigma}{\Theta_\mathrm{HPBW}(\nu)}) ^2
\end{align}
where $\sigma$ is the pointing rms error of FAST which is $8^{\prime\prime}$, we use the beam size($\Theta_\mathrm{HPBW}(\nu)$) of FAST in Table 2 of \citet{jiang2020fundamental}. The value of $f_G$ is 1.0079 at 1100 MHz and 1.0095 at 1400 MHz.

\subsection{Observation data}
The calibrator's observational data serves two research purposes: evaluating various calibration methods and examining gain stability over extended periods. For the first part, we performed continuous observations of 3C48 using three different observation modes to assess the flux density calibration accuracy under varying modes. These observations were conducted on September 9th, 2021, and June 29th, 2022, respectively. During the September 9th, 2021, observations, we utilized the MultiBeamCalibration, DriftWithAngle, and OnOff modes, while during the June 29th, 2022, observations, we replaced the DriftWithAngle mode with the MultiBeamOTF mode to calibrate all beams of FAST. For the second part, the long-term calibrator observations for the M31 halo \HI survey were carried out subsequent to each DriftWithAngle survey observation. We performed flux calibration for each beam roughly every 10 days, accumulating the calibration data over a period of more than 175 days (for the center beam, also M01), from 2020 September to 2022 September. Additionally, a limited amount of publicly accessible 3C286 data was employed to enhance our study when 3C48 was not proper for observations, with these observations being carried out in MultibeamCalibration mode.

\section{data reduction}
\label{sect:data}

All data used in this paper are processed with our pipeline \HIFAST \citep{2024SCPMA..6759514J}, including frequency-dependent noise diode calibration, baseline fitting, standing wave removal, flux density calibration, stray radiation correction, and gridding to produce data cubes. For further details, please consult the \HIFAST series paper. This section is dedicated solely to the model for flux density calibration.

The \HIFASTGAIN module of the \HIFAST pipeline is used to calibrate the flux density gain of FAST. First, the observed power of the calibrator ($P_\mathrm{obs}$) is converted to the antenna temperature ($T_\mathrm{a}$) by using the noise diode signal with a fixed known temperature of $T_\mathrm{cal}$ \footnote{\url{https://fast.bao.ac.cn/cms/category/telescope_performence_en/noise_diode_calibration_report_en/}}:

\begin{align}
T_\mathrm{a}(\nu)= \frac{{P_\mathrm{obs}(\nu)}}{P_{\mathrm{cal}}(\nu)} \times T_{\mathrm{cal}}(\nu), \label{eq:s_p_Tcal},
\end{align}
where $P_{cal}$ is the recorded power of the injected noise diode.

With $T_\mathrm{a}$ obtained for each spectrum per unit frequency, for a source with given flux $S_c(\nu)$, $G$ can be expressed as 
\begin{equation}  
    G(\nu) = \frac{T_\mathrm{a}^\mathrm{SRC}(\nu)-T_\mathrm{a}^\mathrm{REF}(\nu)}{S_c(\nu)}
\label{eq: computeG}
\end{equation}
where $T_\mathrm{a} ^\mathrm{SRC}$ is the antenna temperature of the spectrum in the source area, $T_\mathrm{a} ^\mathrm{REF}$  is the spectrum in the reference area. We briefly discuss the techniques to obtain $T_\mathrm{a} ^\mathrm{SRC}$ and $T_\mathrm{a} ^\mathrm{REF}$  separately in the following two subsections.

\subsection{Tracking modes}
\label{sect:datatrack}

When using tracking modes such as OnOff and MultiBeamCalibration, the FAST backend records the sampling time of the spectrum, which is divided into source (SRC) and reference (REF) groups. The mean spectrum in each group is used to calculate the antenna temperature of the source and reference ($T_\mathrm{a} ^\mathrm{SRC}$ and $T_\mathrm{a} ^\mathrm{REF}$). By introducing these two values into Equation~\ref{eq: computeG}, the gain can be determined. 

\begin{figure}
\centering
\includegraphics[width=0.5\textwidth, angle=0]{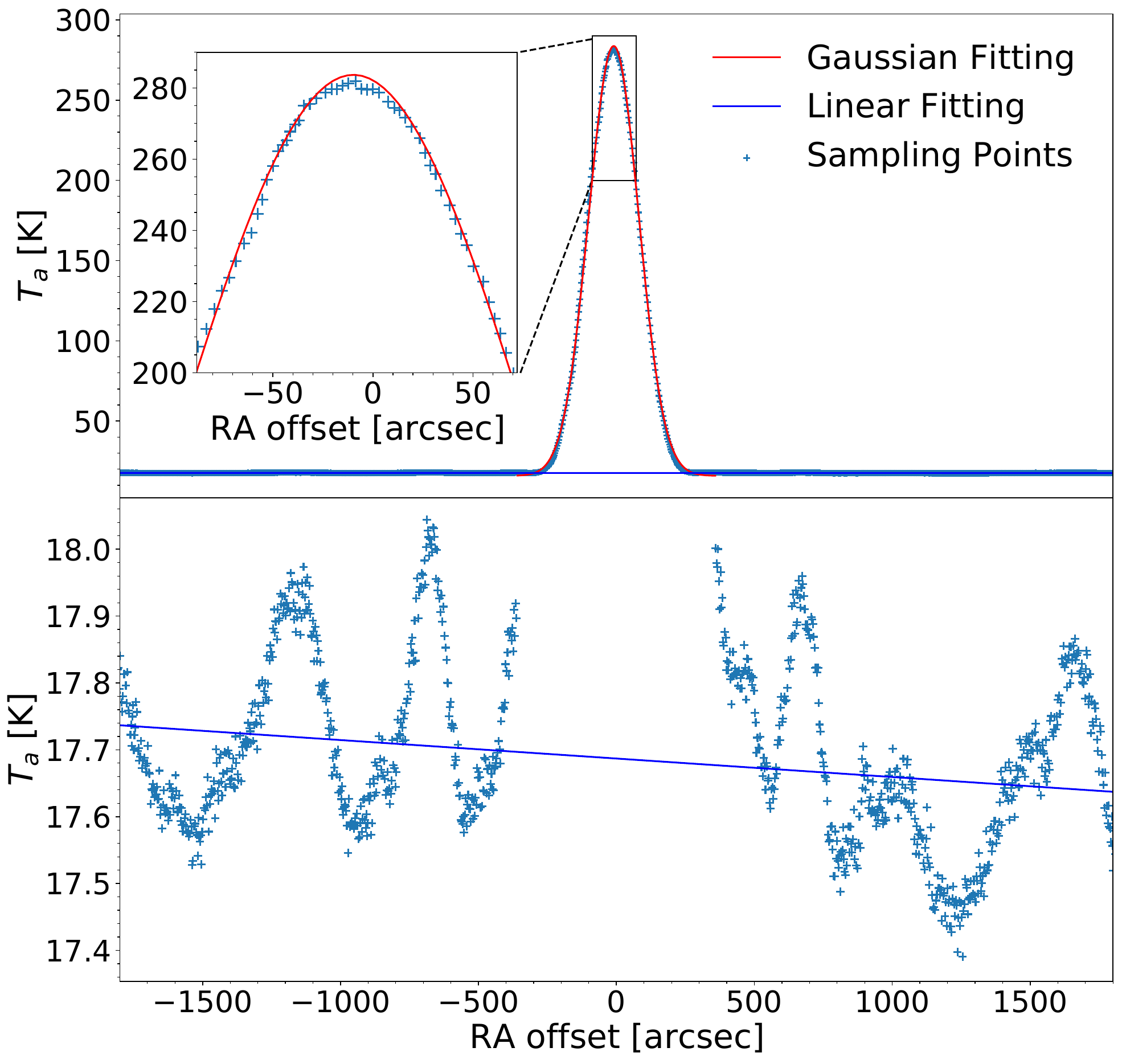}
\caption{An example of Gaussian function fitting results of DriftWithAngle observation. The blue dots represent the sample points, and the red line is the best-fit curve. The $x$-axis is the arc second offset between the sample point and the calibrator, and the $y$-axis is $T_\mathrm{a}$. The top panel is fitted with a Gaussian function, and the bottom panel is fitted with a linear function.}
\label{FigDriftobs}
\end{figure}

\subsection{Scanning modes}
\label{datascan}

In scanning modes such as DriftWithAngle and MultiBeamOTF, the power of the observed spectra will vary significantly as the beam passes over the source. This is the result of the convolution of the calibrator source and the shape of the beam. Generally, the main lobe of the FAST beams can be approximated as a Gaussian function \citep{jiang2020fundamental,xi2022fast}. To obtain $T_\mathrm{a} ^\mathrm{SRC}$ and $T_\mathrm{a} ^\mathrm{REF}$, we use the following function to fit the observed data:

\begin{equation}
    f(d,\nu) = T_\mathrm{a}^\mathrm{SRC}(d,\nu) \times e^{ - \frac{(d-d_0)^2}{2\sigma(\nu) ^2)}} + T_\mathrm{a}^\mathrm{REF}(d,\nu)
\end{equation}
where $d$ is the angular distance offset, $d_0$ is the position of calibrator position, $\sigma$ is the standard deviation depended on frequency. 

The upper panel of Figure \ref{FigDriftobs} displays an example of a Gaussian fit to the temperature variance of Beam 1. When the total residual of the fitting reaches its minimum, $T_\mathrm{a} ^\mathrm{SRC}$ and $T_\mathrm{a} ^\mathrm{REF}$ can be determined simultaneously.
Assuming the main lobe shape follows a Gaussian function may introduce an error of approximately 3\%, given its significantly more complex nature. To reduce this error, we fit separately for $T_\mathrm{a} ^\mathrm{SRC}$ and $T_\mathrm{a} ^\mathrm{REF}$. For $T_\mathrm{a} ^\mathrm{REF}$, a linear function is used to fit the reference area, as illustrated in the lower part of Figure \ref{FigDriftobs}. Then, the equation is applied to the data points passing through the calibration source to determine $T_\mathrm{a} ^\mathrm{SRC}$. With $T_\mathrm{a} ^\mathrm{REF}$ and $T_\mathrm{a} ^\mathrm{REF}$, we then obtain G with Eq. \ref{eq: computeG}. 

\begin{figure}
\centering
\includegraphics[width=0.5\textwidth, angle=0]{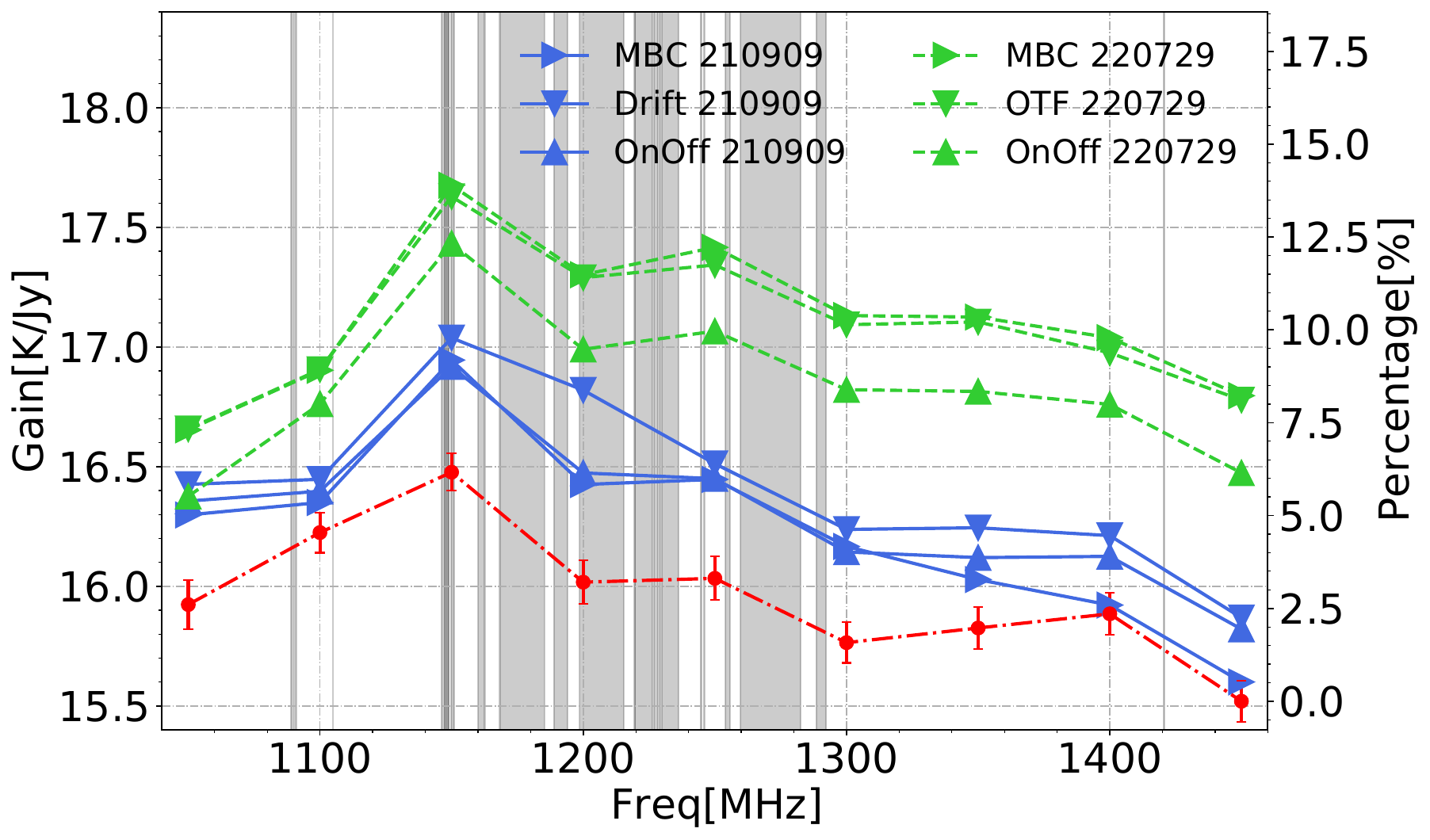}
\caption{Gain parameter $G$ of beam 1 among different methods. Symbols are for different modes respectively. The results from two different days (2021 September 9, 2021, and 2022 July 29) are presented by solid blue lines and dashed green lines respectively. The red dashed line and error bar are the results from $G_{J2020}$ which is observed with the OnOff mode. The gray-shadowed region shows the high probability frequency band ($>50\%$) in \citet{zhang2022radio}.}
\label{FigGainMode}
\end{figure}

\section{results}
\label{sect:randd}

\subsection{Distinction of gain with different observation modes}
\label{sect:diffinmodes}

To investigate the influence of observation modes on calibration accuracy, we conducted continuous observations of 3C48 using three modes on two different dates: September 9th, 2021 (MultiBeamCalibration, DriftWithAngle, and OnOff) and June 29th, 2022 (replacing DriftWithAngle with MultiBeamOTF to obtain $G$ for all beams). The electronic gain fluctuations of the FAST system are known to be better than 1\% over hours of observation \citep{jiang2020fundamental}. Additionally, the aperture efficiency of FAST varies as a function of $\theta_{ZA}$, we corrected for the influence of $\theta_{ZA}$ in processing the flux density gain data using the fitting results of \citet{jiang2020fundamental}. Therefore, the observation mode will not significantly affect the value of $G$ if the difference in $G$ obtained on the same date is in a similar order.

Figure \ref{FigGainMode} illustrates the variation in $G$ for Beam 1 across three observation modes and two distinct days. The left and right y-axes represent the gain parameter and the difference in ratio, respectively. Additionally, for comparison purposes, the results of $G_{J2020}$ (\cite{jiang2020fundamental}) are depicted as red dotted-dashed lines. Different line styles (solid and dashed) denote observations conducted on different dates, while distinct line colours differentiate between various observation modes.

First, the disparity between the various observation modes is quite minimal, with a difference of less than two percent across the entire frequency range. Secondly, The MultiBeamOTF observation on 2021 September 9 at 1200 MHz yielded a higher $G$ than the other two modes, which is likely due to the presence of strong RFI signals mixed in with the observational signal. In the shadow range, strong RFI is likely caused by communication and navigation satellites \citep{zhang2022radio, wang2021satellite}. So the measurement of the response of FAST received at $1000 - 1500$ MHz does not depend on the measured method, but could be contaminated by the strong RFI. Thirdly, $G$ has a large variation with the observing date. For instance, the measurement of $G$ on 2022 June 29 exhibited an increase of nearly 5\% compared to the measurement on 2021 September 9, and a 7\% increase compared to the value reported in $G_{J2020}$. This indicates that the $G$ parameter seems to vary on longer timescales of days or even longer.

\subsection{Variation of gain in long-term observation}
\label{sect:Observationdate}

\begin{figure*}
\centering
\includegraphics[width=\textwidth, angle=0]{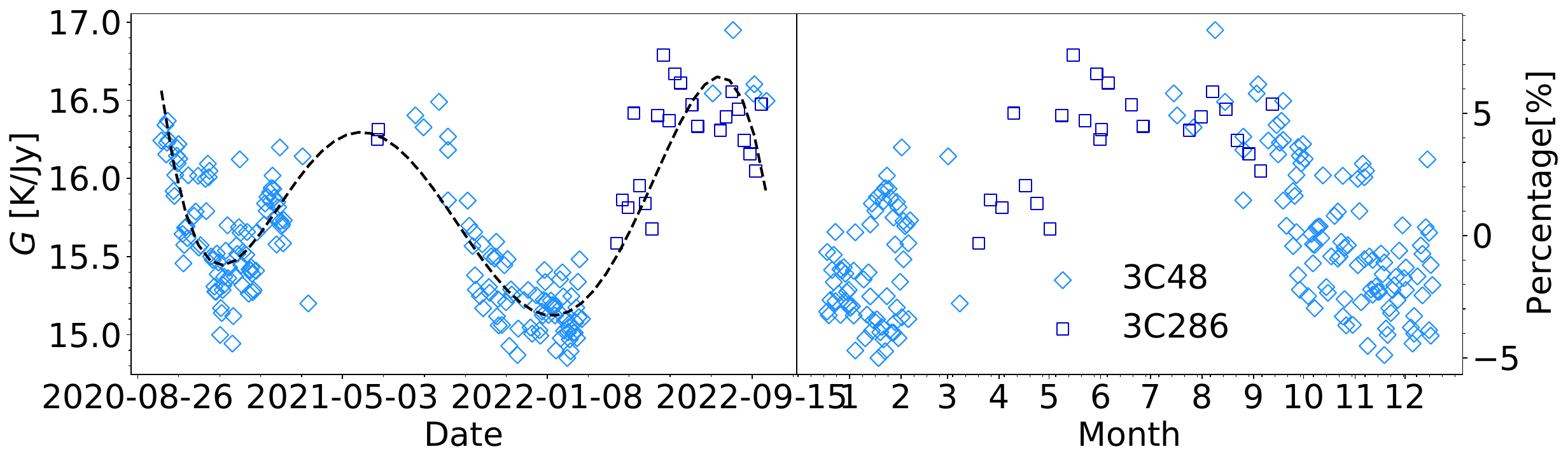}
\caption{$left$: Gain parameter $G$ of Beam 1 at 1400 MHz from 2020 September to 2022 September. The diamond marker is $G$ of 3C48 and the square marker is 3C286. The black dashed line represents the temporal variation of $G$ fitted by polynomial regression. $right$: $G$ in the left panel rearranged by months, $y$--axis is the difference between $G$ with mean value in percentage.}
\label{fig:gain_time}
\end{figure*}

To better understand the long-term fluctuations of $G$, we conducted tests on Beam 1 at around 1400 MHz (averaged with a frequency width of 1 MHz) and plotted the results in the left panel of Figure \ref{fig:gain_time}. The $x$--axis represents the observation time, while the $y$--axis represents the value of $G$. We used 3C48 as the main calibrator because its coordinates are close to our survey area. However, most calibrations in May and June were performed with 3C286, as 3C48 was in the Solar avoidance zone.

We find that the maximum fluctuation of $G$ is around 8\% in 2022 September compared to the mean value of $G$. Our observations show that the values of $G$ are generally lower in winter and higher in late summer, as shown in the right panel of Figure~\ref{fig:gain_time}, we arranged the data by month. This suggests that, in addition to system hardware configuration changes, the FAST telescope's response may be influenced by yearly periodical factors, such as temperature and air pressure. Next, we roughly obtained the relationship between $G$ and observation time using polynomial regression, as shown in Figure~\ref{fig:gain_time}. The polynomial is presented below:

\begin{equation}
     G(t) = a_5 t^5 + a_4 t^4 + a_3 t^3 + a_2 t^2 + a_1 t + a_0
\label{eq:gain-time}
\end{equation}
where $t= MJD_{obs}-59115$ is MJD time of observation, the parameters of the polynomial are $a_5=-1.242\times10^{-12}, a_4=2.298\times10^{-9}, a_3=-1.481\times10^{-6}, a_2 =3.897\times10^{-4}, a_1 t=-3.710\times10^{-2}, a_0=16.59$.

\begin{figure}
\centering
\includegraphics[width=0.5\textwidth,angle=0]{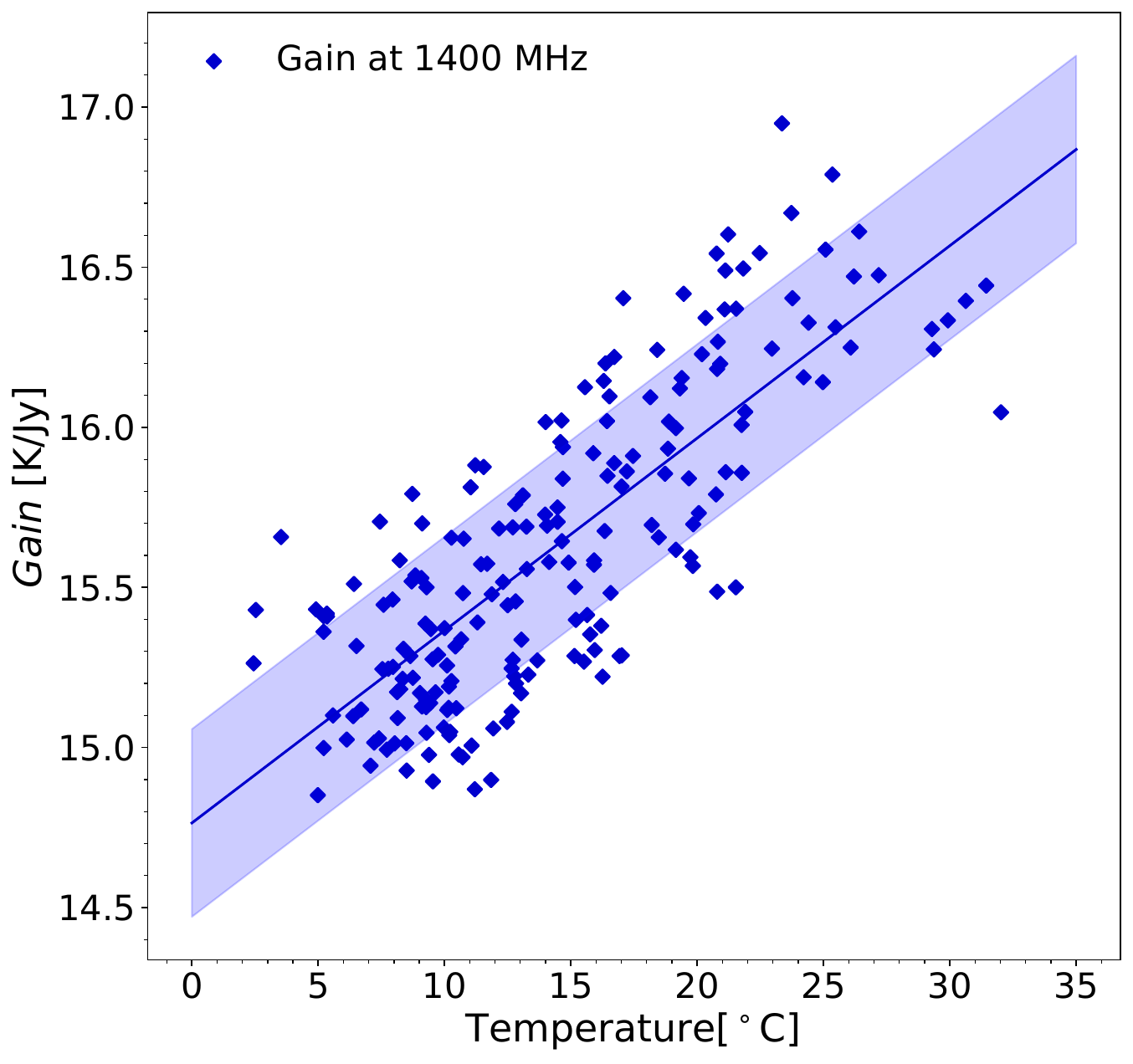}
\caption{ The correlation between $G$ and atmospheric temperature. The $x$--axis represents temperature and the $y$--axis represents the $G$. The data points represent the $G$ of the calibrators at 1400 MHz, and the blue line represents the correlation curve fitted using the least squares method.}
\label{G-T}
\end{figure}

\begin{figure}
\centering
\includegraphics[width=0.5\textwidth,angle=0]{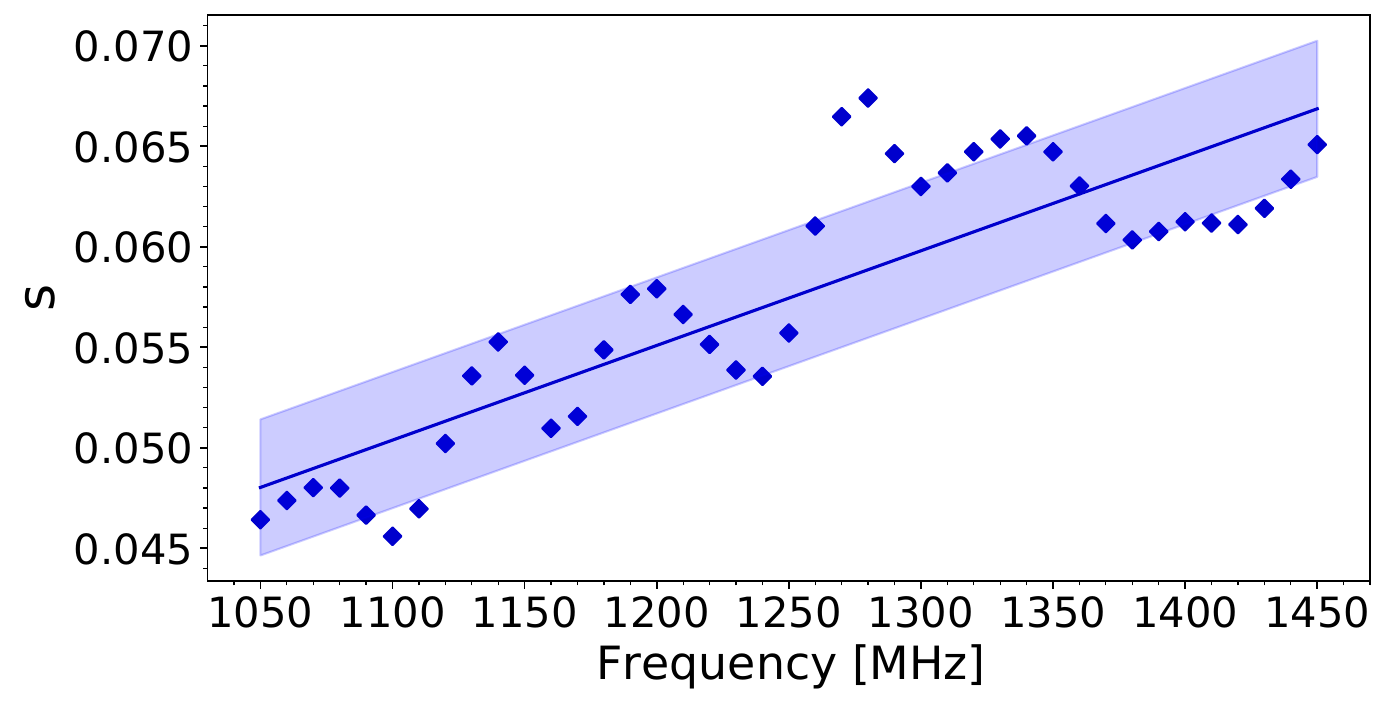}
\caption{ The relationship between the frequency and slope of the G--T relationship. The $x$--axis represents frequency values, and the $y$--axis represents the values of the slope. Although there are some fluctuations in the variation of the slope, overall it still exhibits a linear change with frequency.}
\label{GTslope}
\end{figure}

To account for the seasonality of the parameter $G$, we examined the atmospheric temperature, pressure, and humidity during the calibration source observations. The results indicate that the variable $G$ is not sensitive to changes in pressure and humidity, while $G$ exhibits a significant relationship with atmospheric temperature. Figure \ref{G-T} shows $G$ as a function of atmospheric temperature $T$, at the observing frequency of 1400 MHz. We performed a least square linear regression between $G$ and atmospheric temperature, with its slope linearly correlated with observing frequency, as depicted in Figure \ref{GTslope}. Both relationships are represented by:

\begin{align}
G_\mathrm{cor}(\nu)= G(\nu) + s(\nu) \Delta T
\label{eq:G-T}
\end{align}

where $s(\nu)=4.7\times10^{-5}\nu-1.4\times 10^{-3}$, $\nu$ represents the frequency of $G$, and $\Delta T$ denotes the temperature difference between the observed value and the standard value.

\begin{figure}
\centering
\includegraphics[width=0.5\textwidth,angle=0]{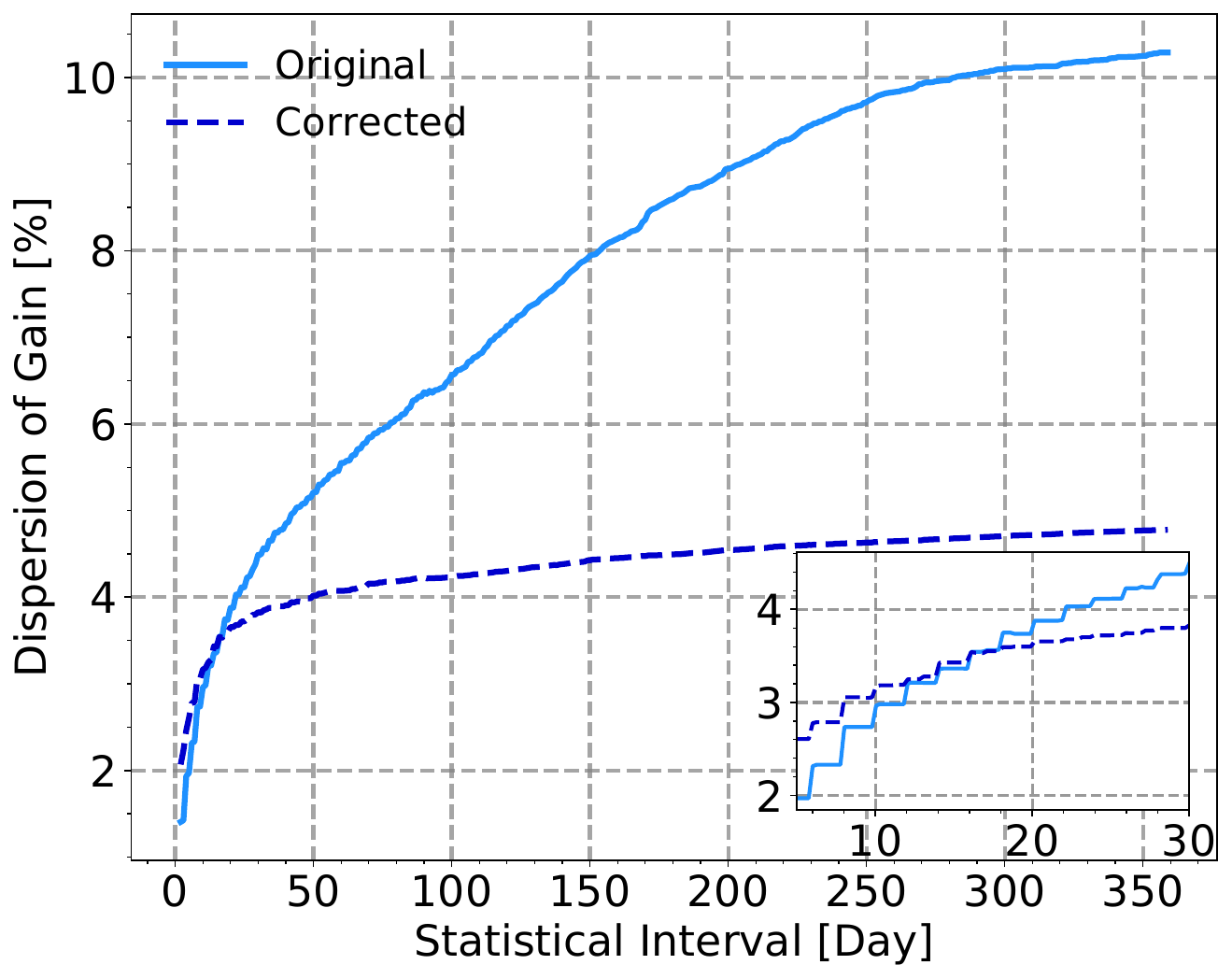}
\caption{ Dispersion of $G$ for different calibrator observation intervals at 1400 MHz. The dispersion of G represents the average variance of $G$ within a fixed interval when traversing all calibrator data using a rolling window approach with different observation intervals. The solid line represents the $3\sigma$ dispersion of the original $G$ value for each interval in percentage, while the dashed line indicates the $G$ corrected by the $G-T$ relationship. The inset plot at the bottom right shows the dispersion of $G$ within the calibration interval of 30 days.}
\label{DispG}
\end{figure}

After correcting the dependence of the gain $G$ on atmospheric temperature, we performed a statistical analysis of the variance of $G$ over a full calendar year.By setting an observation interval window ranging from 1 day to 365 days, and using a rolling window to traverse all data results, we can calculate the average variance of $G$ over this time interval. This method enables us to determine the magnitude of the variation in gain values for different observation time intervals. Figure \ref{DispG} displays the dispersion of $G$ for different time intervals. The solid line represents the $3\sigma$ dispersion of the original $G$ values for each interval, while the dashed line represents $G$ after correction using Equation~\ref{eq:G-T}. Overall, the dispersion of $G$ increases rapidly with longer time intervals. The dispersion of the original $G$ exceeds 10\% when the time interval is longer than 300 days, whereas the corrected $G$ eventually stabilizes and remains below 5\%, even for intervals close to one year. In summary, the gain $G$ exhibits remarkable stability, considering the systematic errors from the telescope and receiver, as well as those introduced during the flux calibration process, which are accounted for in the scatter statistics.

To examine the variability of the scatter for shorter time intervals, a zoomed-in version of the dispersion values of $G$ for intervals less than 30 days is shown in the inset plot at the bottom right of Figure \ref{DispG}. The correction of the G--T relationship does not have a significant impact within these shorter time intervals. The dispersion of $G$ is approximately 3\% over a 10-day time interval in both situations. This suggests that, in the absence of any calibrator observations, the flux calibration process could introduce an error of up to 10\%. However, this error can be reduced to 3\% if at least one calibrator observation is carried out within 10 days. Moreover, if calibrator observations are performed on a daily routine, the error can be further reduced to within 2\%. On the other hand, if there is a long time lag between the calibrator data and the target field data, it becomes necessary to correct the gain using the G--T relationship.

\subsection{Gain curve variation among the multiple beams}
\label{sect:Frequency}

\begin{figure}
\centering
\includegraphics[width=0.5\textwidth, angle=0]{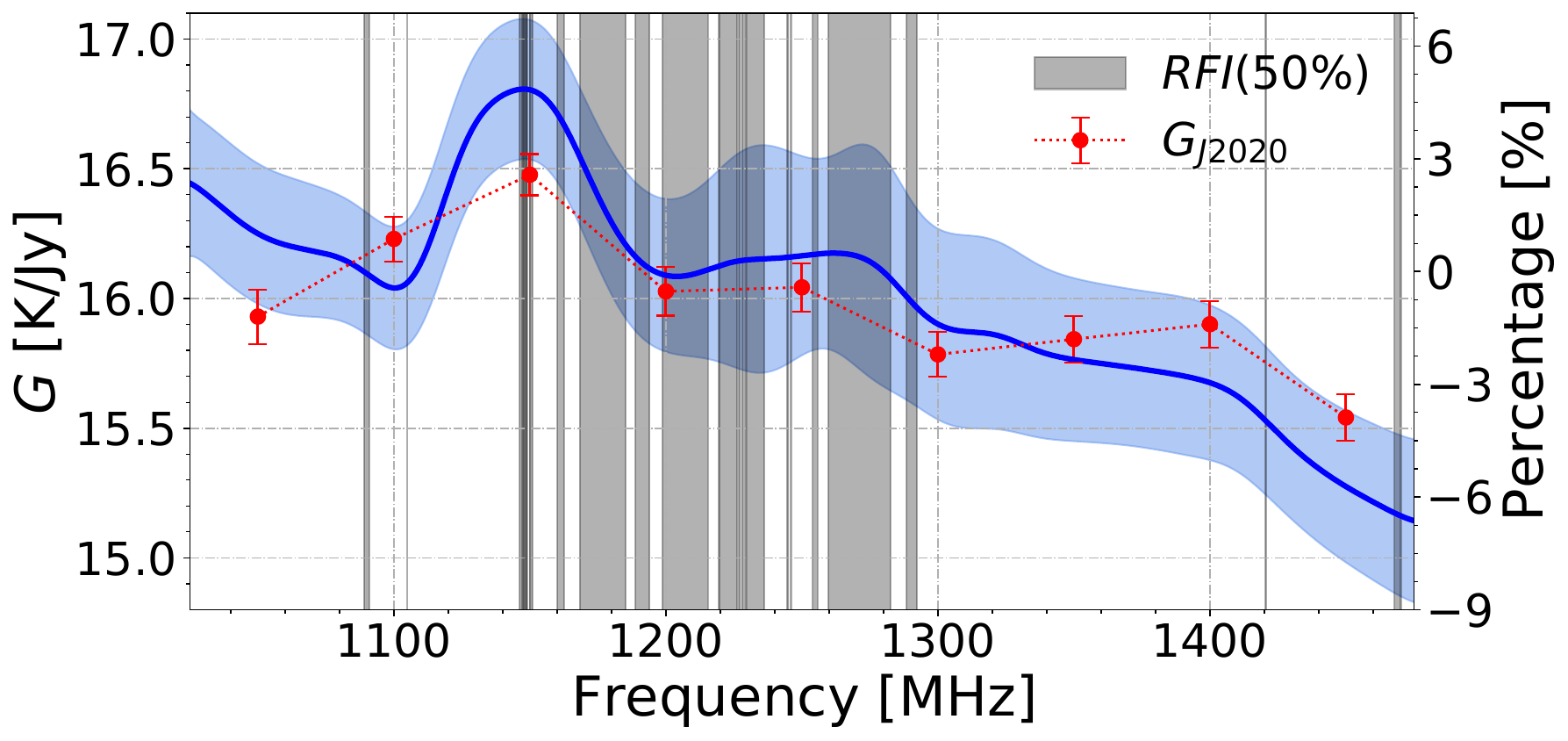}
\caption{Frequency dependence of $G$ of Beam 1. The blue line is the mean value of flux density gain in all calibration data corrected from the G--T relationship, the pale region shows the $1\sigma$ range of corrected--$G$ fluctuation. The red dot and error bar is $G_{J2020}$. The right $y$--axis shows the difference of $G$ in percentage. The gray-shadowed region shows the high probability frequency band ($>$50\%) in \citet{zhang2022radio}. }
\label{fig_gain_freq}
\end{figure}

During our \HI deep survey around the M31 halo region, we performed flux calibration for each beam roughly every 10 days, accumulating the calibration data over a period of more than 175 days (for the center beam, also M01), from 2020 September to 2022 September. Leveraging these  observations, we were able to determine the frequency dependency of the $G$ (referred to as the gain curve) for every beam. Figure \ref{fig_gain_freq} shows the gain curve for M01. The solid line represents the mean value of $G$ at each frequency channel, while the blue shadowed region signifies the dispersion of $G$ post-adjustment. The results for $G_{J2020}$ are represented by red symbols with error bars. In general, the gain parameter gradually declines as the frequency surges, except for a small hump around a frequency of 1150 MHz. Our findings align with $G_{J2020}$ within the 1 sigma error of the data adjusted by $G-T$ relation. Except for a single data point, again at a frequency of 1150 MHz. This aberration could be ascribed to considerable RFI contamination. The likelihood of RFI emerging around this frequency point is roughly 80\% and 50\% for the XX and YY channels, respectively, as indicated by \citep{zhang2022radio}. 

\begin{figure}
\centering
\includegraphics[width=0.5\textwidth,angle=0]{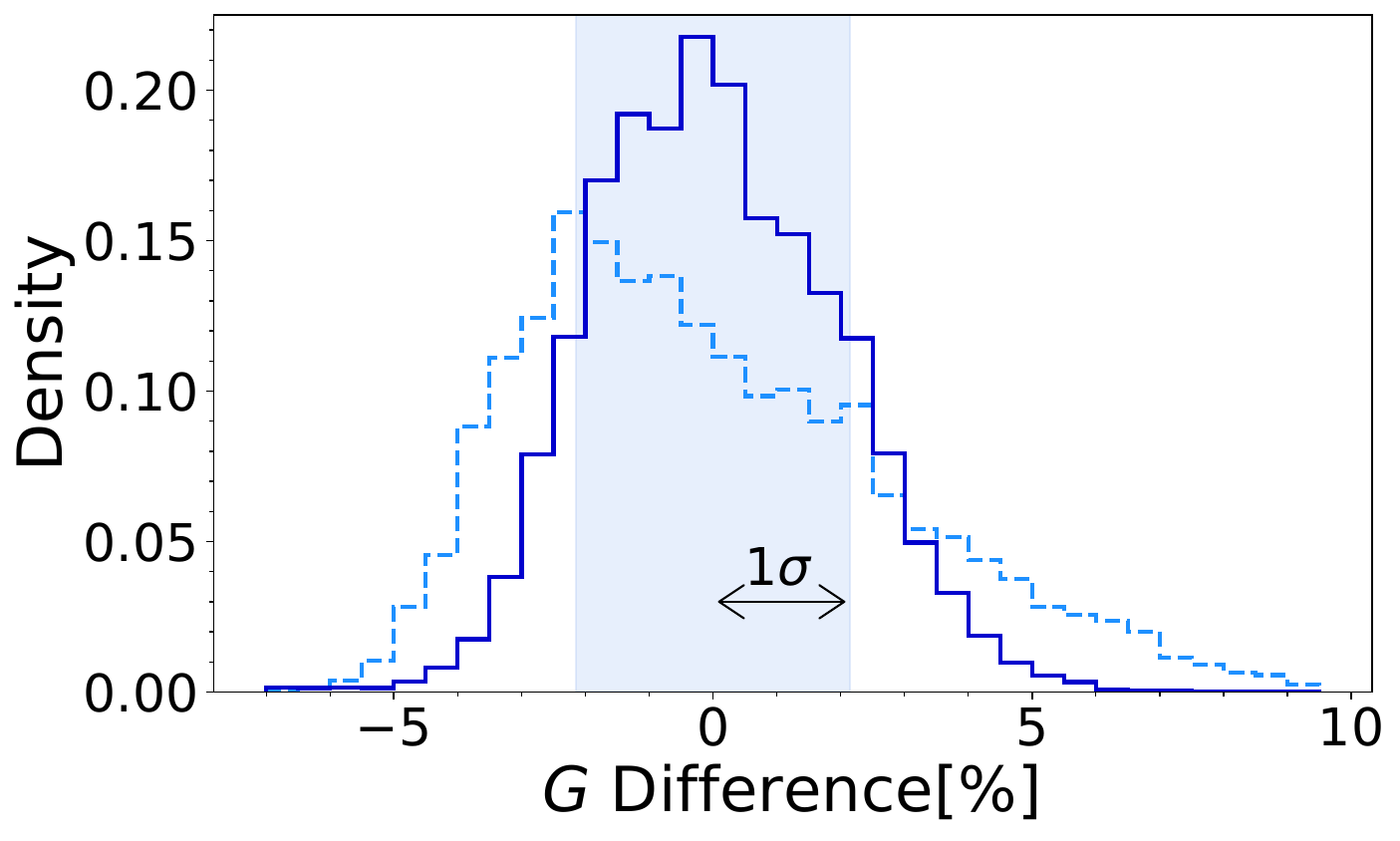}
\caption{The probability distribution function (PDF) of parameter $G$ normalized by the mean value around 1400 Mhz. The dashed blue line collected original data from every calibration, while the solid line shows data corrected by the $G-T$ relationship. The $1\sigma$ ranges for the two cases are 3.2\% and 2.15\% respectively.}
\label{gaindiffpdf}
\end{figure}

Furthermore, in Figure \ref{gaindiffpdf}, we elucidate the probability density function (PDF) of the fractional difference of the $G$ parameter around 1400 MHz normalized by the average value, in percentage terms. The results with and without G-T adjustments are denoted by solid and dashed histograms, respectively. For the corrected results, one sigma is merely about 2.15\%, denoted by the shadowed area. The G-T adjustment considerably refines the probability distribution of the G parameter, reorienting the original data skewed towards lower values (owing to most of the observation data being collected in the colder seasons, causing an overall dip below the actual annual average) more towards the central sector. 

\begin{figure*}
\centering
\includegraphics[width=13cm,angle=0]{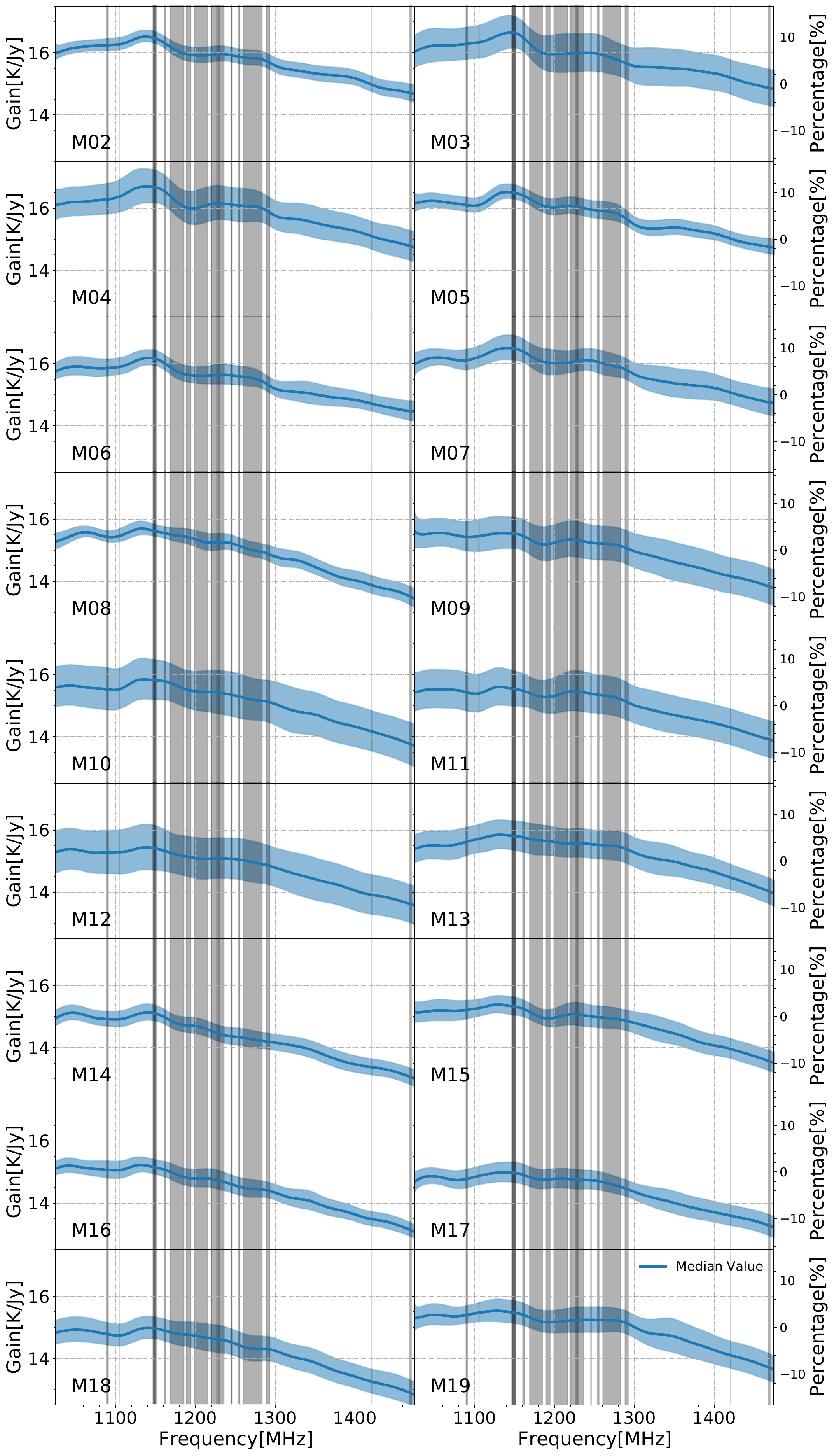}
\caption{Corrected Gain parameter by G-T relation for outer circle beams (Beam 2 to Beam 19). The blue shadowed region indicated 1 $\sigma$ range. The gray shadows correspond to the same region as in the previous figure}
\label{Gain2to19_corrected}
\end{figure*}

The gain curves for the remaining 18 beams are presented in Figure \ref{Gain2to19_corrected} (with G--T adjustment), the blue shadowed region is 1$\sigma$ dispersion of $G$. For the full-beam calibration observations, a total of 17 sessions were conducted, while observations for M02, M05, M08, and M14 were carried out 84 times in total (utilizing the DriftWithAngle mode as described in Sec \ref{sect:observationmode}, allowing calibration of five beams in a single observation). Consequently, the data dispersion for these four beams is relatively low. Bands heavily contaminated by strong RFI are shaded in gray. 

\section{discussions}
\label{sect: disc}

\begin{figure}
\centering
\includegraphics[width=0.5\textwidth,angle=0]{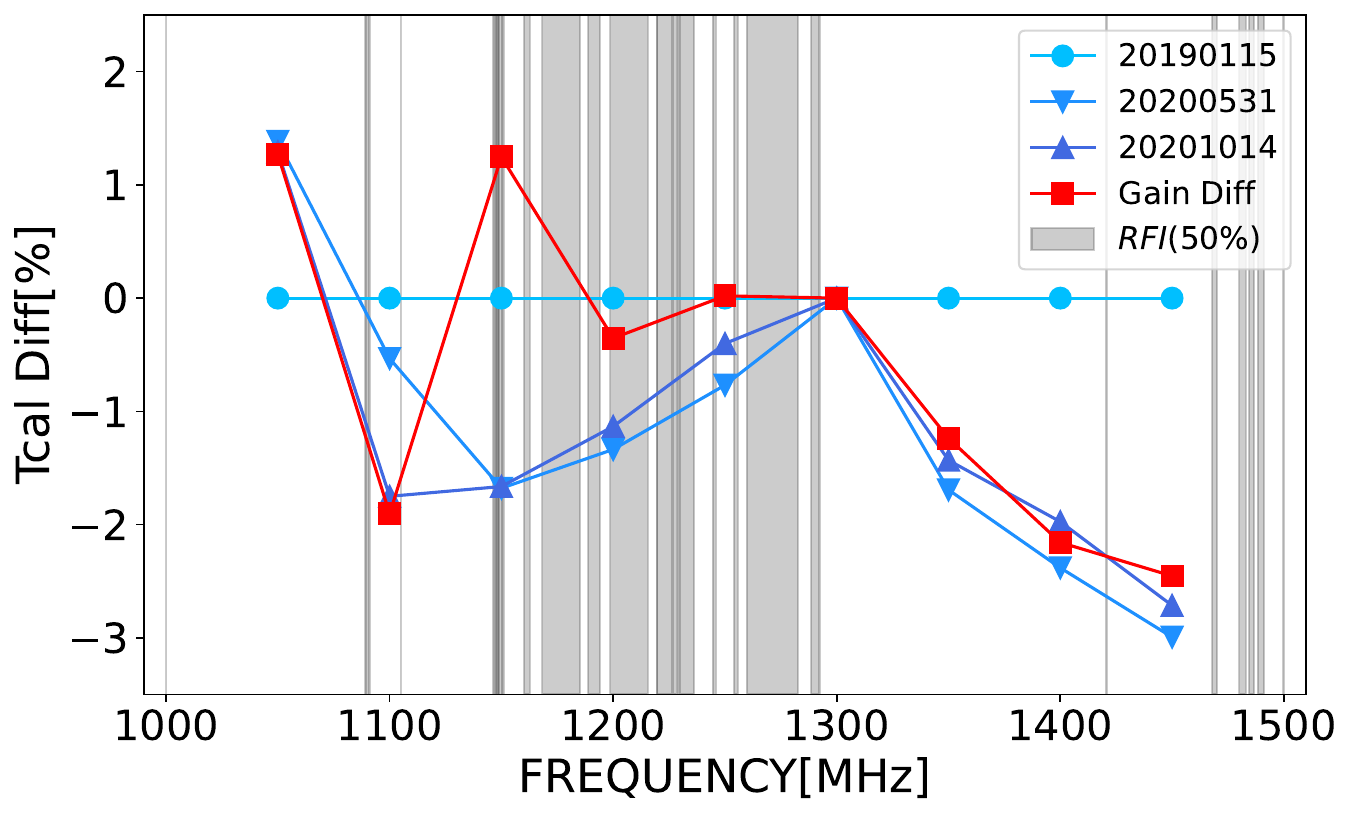}
\caption{$T_{cal}$ difference of Beam 1 at a frequency higher than 1300 MHz, different lines show the percentage of the difference between each $T_{cal}$ file and $T_{cal}$ on 2019 January 15. The $T_{cal}$ is smoothed with the frequency range of 50 MHz. The difference of $G$ is shown in a red line with square markers. }
\label{TcalDiff}
\end{figure}

Given the measured $G-T$ relationship, the differences between $G$ and $G_{J2020}$ become intricate. It is further complicated by the fact that the measurements of $G_{J2020}$ for various beams were conducted on different dates. Mismatches exist in both the shape of the gain curve and the values of $G$. Within this section, we first discuss the possible origins of the $G-T$ relationship, then discuss other origins of absolute flux calibration, and then present our online database for the Gain calibration. More nuanced discussions regarding result comparisons are presented in \ref{sec:appdis}.

\subsection{Possible origin of the $G-T$ relation}
\label{sect:diss-G-T}

Various factors can contribute to the fluctuations of $G$, such as changes in the telescope response, instability of the noise diodes, and pointing errors. However, as shown in Section \ref{sect:pointing}, the pointing is unlikely to cause significant errors in calibrator observation. Hence, the origin of the $G-T$ relationship obtained in Section \ref{sect:Observationdate} becomes intriguing. 

\subsection{Hardware update during the observing epochs}
\label{sect:diss-G-T}

Variations in the shape of the gain-curve profiles even go beyond the adjustments made through the $G-T$ relationship correction above in Figure~\ref{fig_gain_freq}. We find that these discrepancies could be a result of inaccuracies introduced by the noise diode files during the calibration process of $T_{cal}$ as discussed in Section \ref{sect:data}. Differences in the shape of the gain curve may be due to the different conversion coefficients of the noise diodes ($T_{cal}$) used in Equation~\ref{eq:s_p_Tcal}. The variations in $T_{cal}$ among three different versions (20190115, 20200531, and 20201014 are illustrated in Figure \ref{TcalDiff}. The first version is used to determine $G_{J2020}$ and the others are used in our data processing. We observed that the differences in $T_{cal}$ (shown by the blue lines) closely correspond to the differences in $G$ indicated by the red line in Figure \ref{TcalDiff}, in most frequency ranges, except at 1150 MHz, where serious interference from RFI is present. Taking into account the numerous sources of error in the flux calibration process, we suggest that the disparities between $G_{J2020}$ and our calculated gain curve primarily arise from the adjustments made through the G--T relationship correction and the calibration of the noise diode, while also being influenced by the presence of irregular RFI signals. 

\subsection{Temperature-sensitive telescope responses}
\label{sect:diss-G-T}

However, the influence exerted by the $T_{cal}$ files tends to remain consistent over a given period and does not typically manifest as periodic effects. Notably, the relative deviations observed in the more recent two versions of $T_{cal}$ files depicted in Figure~\ref{fig_gain_freq} consistently exhibit the same directional bias. Therefore, it is improbable that they would result in scenarios where one aspect of $G$ increases while another diminishes. Consequently, we hypothesize that the primary source of the $G-T$ relationship may derive more prominently from the inherent response characteristics of the telescope itself. It is challenging to pinpoint precisely which part of the telescope response is affected by this influence. Given the stringent correlation between $G$ and atmospheric temperature, we speculate that temperature-induced thermal expansion and contraction of certain hardware structures within the telescope might introduce interference during observations, such as defocus or reduced panel reflection efficiency, ultimately manifesting as periodic fluctuations in $G$. Detailed hardware examination is beyond the scope of this work. 

\subsection{Intrinsic flux variation from flux calibrators}
\label{sect:diss-G-T}

Additionally, according to the flux density calibration guidance for the VLA\footnote{\url{https://science.nrao.edu/facilities/vla/docs/manuals/oss/performance/fdscale}}, the calibrator 3C48 has been undergoing a flare since 2018 or so. Other instruments beyond VLA have indicated measurements of the flare's impact intensity at certain frequencies, with the L-band showing a relatively minor 5\% influence\citep{vla2021}. The flux variations induced by the flares of 3C48 may also be reflected in long-term observations of $G$. The blue dashed line in Figure~\ref{DispG} represents the discretization of the $G$ values over extended observations after eliminating the influence of the $G-T$ relationship, with this value tending towards $\sim 5\%$. However, this discretization should already encompass both the fluctuations in the flux density of 3C48 itself and the systematic gain errors during FAST observations. Thus, we infer that the impact of the flares of 3C48 on flux density calibration is expected to be less than 5\% and unlikely to significantly affect our other findings.

\subsection{Guideline of utilizing the detailed data}
Detailed data on the $G$ values for 19 beams covering the frequency range 1000 MHz -- 1500 MHz can be found on the \HIFAST homepage: \url{https://hifast.readthedocs.io/fluxgain}. 

These data are suitable for direct application in flux density calibration of observation targets with a zenith angle near $10.15^{\circ}$ and an atmospheric temperature close to 15 degrees Celsius. When the zenith angle differs from this orientation, conversion using the data and formulas provided in \citet{jiang2020fundamental} is necessary. Furthermore, significant deviations from the standard temperature require adjustment for the $G-T$ relationship using Equation~\ref{eq:G-T}. Access to atmospheric temperature data during observations can be obtained by contacting the FAST team. Note that Equation~\ref{eq:gain-time}, fitted to obtain $G$, is applicable only for observations requiring flux density calibration conducted during the time depicted in Figure~\ref{fig:gain_time}. 

\section{conclusions}
\label{sect:Conclusions}

In this study, we conducted a comprehensive investigation into the performance of the flux density gain parameters ($G$) of FAST in different observation modes and checked the variation of gain in long-term observations. Our findings provide valuable insights into the error range of parameter $G$ in data processing. We find that the differences in $G$ obtained by different observation modes are negligible.
Also, the variation of the gain parameter is small even over a time scale of years, although the presence of strong RFI will reduce the reliability of $G$ significantly, e.g. in the frequency range from 1150 MHz to 1300 MHz. 

Furthermore, our analysis of the long-term fluctuation of $G$ revealed a linear correlation between $G$ and atmospheric temperature. We showed that correcting for temperature-induced errors in $G$ can substantially reduce the dispersion of $G$ in long-term observations. After this correction, the 1 sigma variance of $G$ could be reduced to less than 5\% for all beams. The gain curves for all 19 beams within the frequency range of 1000--1500 MHz have been provided and are available for download from the \HIFAST homepage. 

Additionally, the test presented in Appendix B behooves us to note that when engaging in data processing, it is imperative to utilize the noise diode files temporally closest to the observation time. Any deviation from this established best practice could potentially induce errors, thus compromising the reliability and validity of our findings.

\begin{acknowledgements}
We would like to thank the referee for their constructive comments.We acknowledge the support of the China National Key Program for Science and Technology Research and Development of China (2022YFA1602901,2023YFA1608204), the National Natural Science Foundation of China (Nos. 11988101, 11873051, 12125302, 12373011, 12041305, 12173016.), the CAS Project for Young Scientists in Basic Research Grant (No. YSBR-062), and the K.C. Wong Education Foundation, and the science research grants from the China Manned Space Project. Y Jing acknowledges support from the Cultivation Project for FAST Scientific Payoff and Research Achievement of CAMS-CAS. 

This work has used the data from the Five-hundred-meter Aperture Spherical radio Telescope ( FAST ). FAST is a Chinese national mega-science facility, operated by the National Astronomical Observatories of the Chinese Academy of Sciences (NAOC).
\end{acknowledgements}

\bibliographystyle{raa}
\bibliography{ms2024-0101}

\begin{appendix}
\renewcommand{\thesection}{Appendix}

\section{comparison with $G_{J2020}$} \label{sec:appdis}

\begin{figure}
\centering
\includegraphics[width=0.5\textwidth,angle=0]{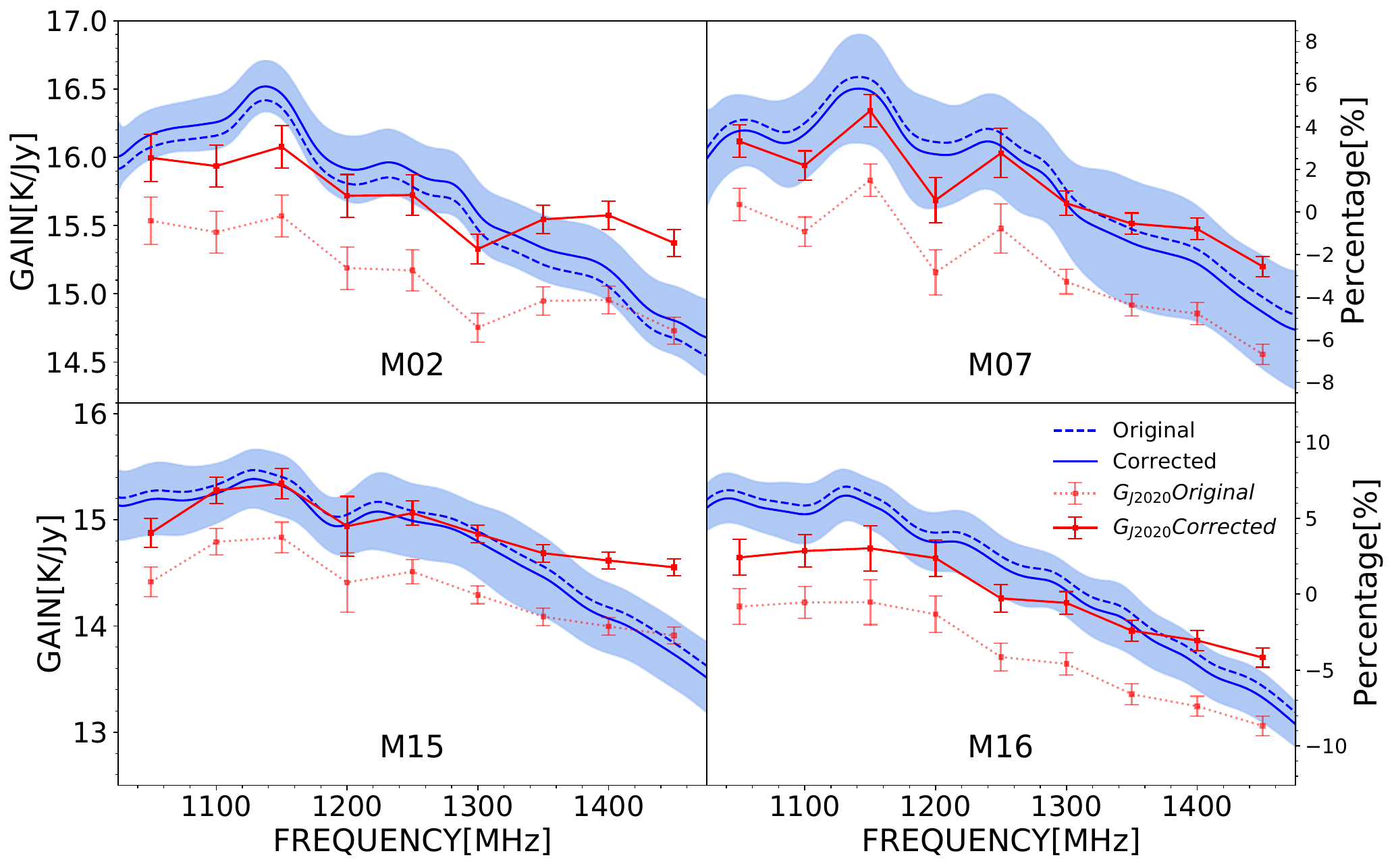}
\caption{Gain parameters for M02, M07, M15, and M16. The original gain curve is represented by dashed lines, while the results corrected by the G--T relationship are illustrated by solid lines. The blue-shaded region represents the 1$\sigma$ area of the $G$ dispersion after the $G-T$ adjustment.}
\label{Gain271516}
\end{figure}

\begin{figure*}
\centering
\includegraphics[width=13cm,angle=0]{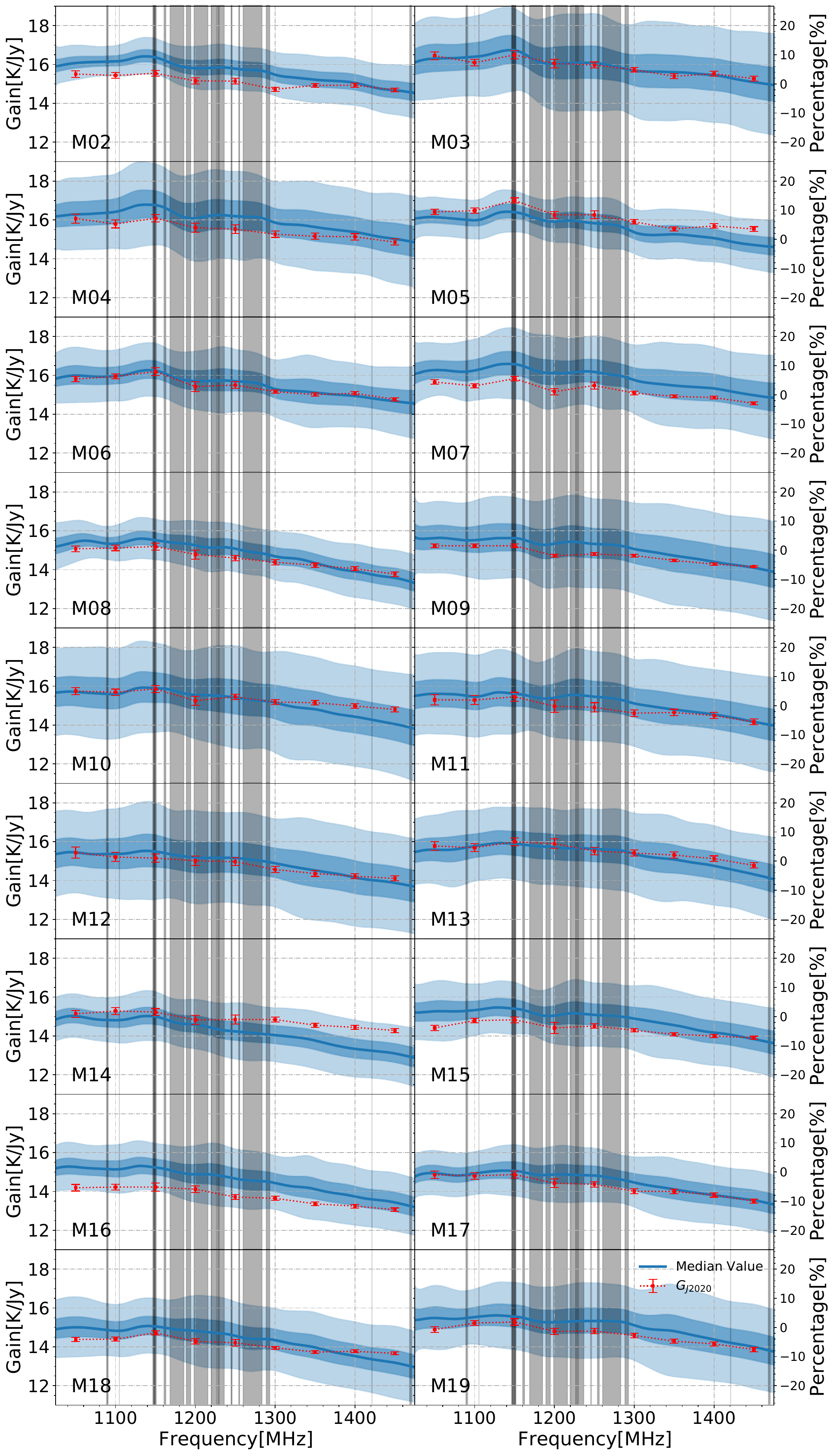}
\caption{Frequency dependence of the Gain parameter for outer circle beams (Beam 2 to Beam 19). The pale blue and grey blue shadowed region indicate 1 and 3 $\sigma$ range for the dispersion of original data separately. The red dots and err bar are $G_{J2020}$. The right $y$-axis shows the difference of $G$ in percentage. The gray shadowed region shows the frequency bands that have high probability ($>$50 \%)  covered by RFI.}
\label{Gain2to19}
\end{figure*}

In this section, we delve into a detailed comparative analysis with G2020, focusing on the implications of correcting the G--T relationship on beams surrounding the central beam. Figure \ref{Gain2to19} illustrates the gain curve of the uncorrected peripheral beams, with the $1\sigma$ and $3\sigma$ ranges shown as shaded blue regions. The red data points, along with the error bars, represent the calculated $G_{J2020}$ values at the corresponding zenith angles. Our findings are mainly consistent with $G_{J2020}$ within one $\sigma$ range for most beams. However, beams 2, 7, 15, and 16 indicate values exceeding $G_{J2020}$ by more than one sigma in certain frequency ranges. It is important to note that the parameters $G$ for different beams based on $G_{J2020}$ are obtained from observations on various dates, making them susceptible to fluctuations in environmental temperature, thereby resulting in significant overall deviations. Figure \ref{Gain271516} presents the rectified data corresponding to the parameters $G$ for beams 2, 7, 15, and 16 after adjustment of the G--T relationship. Subsequently, to this adjustment, our observational results demonstrate a closer alignment with those of $G_{J2020}$, with most observational points falling within the 1$\sigma$ error margin. It should be noted that we cannot get the temperature of the exact time for the observation carried out in $G_{J2020}$, the temperature data from observations taken around the same time in the following year are derived from non-systematic environmental temperature recordings at that specific point in time.

\end{appendix}

\label{lastpage}

\end{document}